\documentclass[aps,amsmath,amssymb,floatfix,twocolumn,prb]{revtex4-1}
\usepackage[]{standalone}
\usepackage{times}
\usepackage{graphicx}
\usepackage{color}
\usepackage{bm}
\usepackage{braket}
\usepackage{mathtools}
\usepackage{mathcomp}
\usepackage[caption=false,justification=raggedright,position=top,singlelinecheck=off]{subfig}
\usepackage{soul}
\usepackage[mathscr]{euscript}

\newcommand{\enbe}{\begin{equation}}
\newcommand{\enee}{\end{equation}}

\begin{document}
\title{Solvable model for quantum criticality between 
Sachdev-Ye-Kitaev liquid and disordered Fermi liquid}

\author{Oguzhan Can}
\email[Corresponding author: ]{ocan@phas.ubc.ca}

\author{Marcel Franz}
\affiliation{Department of Physics and Astronomy and Stewart Blusson Quantum Matter Institute,
University of British Columbia, Vancouver, B.C., V6T 1Z1, Canada}

\date{\today}

\begin{abstract}
We propose a simple solvable variant of the Sachdev-Ye-Kitaev (SYK)
model which displays a quantum phase transition from a
fast-scrambling non-Fermi liquid to disordered Fermi liquid. Like the
canonical SYK model, our variant involves a single species of Majorana
fermions connected by all-to-all  random four-fermion
interactions. The phase transition is driven by  a random two-fermion
term added to the Hamiltonian whose structure is inspired by proposed
solid-state realizations of the SYK model. Analytic expressions for
the saddle point solutions at large number $N$  of fermions are
obtained  and show a characteristic scale-invariant $\sim
|\omega|^{-1/2}$ behavior of the spectral function below the
transition which is replaced by a $\sim |\omega|^{-1/3}$ singularity
exactly at the critical point. These results are confirmed by numerical solutions of the saddle point equations and discussed in the broader context of the field.
\end{abstract}
\maketitle

\section{Introduction}
Sachdev-Ye-Kitaev model\cite{SY1996,Kitaev2015} is an exactly solvable
model of a non-Fermi liquid connected to quantum gravity theories
through the holographic principle\cite{Sachdev2015,Maldacena2016}. 
The SYK model and its variants
\cite{Polchinski2016,Fu2016,Witten2016,Altman2016,Berkooz2016,Bi2017,Murugan2017,Peng2017,Lantagne2018}
show intriguing behaviors as well as unexpected relations to seemingly
unrelated areas of physics, ranging from strongly correlated
fermions\cite{Liu2017,Balents2017,Huang2017,CenkeXu2017a,CenkeXu2018a} to quantum
chaos,\cite{Hosur2016,Gu2016,Chen2017,Krishnan2017} quantum information
theory,\cite{Solano2017,Laflamme2017} random matrix
theory\cite{Xu2016,Verbaar2016,Li2017}, many-body localization\cite{Jian2017} and
wormhole dynamics.\cite{Maldacena:2018} Several proposals for experimental realizations of the SYK model have been given\cite{Danshita2017,Pikulin2017,Alicea2017,Achen2018} which raises prospects for testing these ideas in a laboratory. 

The standard SYK model is controlled by a single dimensionless parameter (the
strength of interactions $J$ divided by temperature $T$) and remains
in the same phase for all its values. An interesting class of models
seeks to modify the SYK Hamiltonian such that it undergoes a
transition to another phase by tuning a control parameter. Banerjee
and Altman\cite{Altman2016} introduced an SYK model with coupling to a set of
``auxiliary fermions'' which exhibits a quantum phase
transition from the non-Fermi liquid SYK phase to a disordered Fermi
liquid as the ratio of the number of SYK fermions to the number of
auxiliary fermions is tuned. Bi {\em et al.} \cite{Bi2017} considered
a model with modified structure of four-fermion coupling constants
which likewise undergoes a phase transition out of the SYK liquid as a
dimensionless parameter characterizing this structure is tuned. These
studies are inherently important because they elucidate the limits of
stability of the SYK liquid and show how it relates to other more
conventional quantum phases of interacting fermions. Understanding
these relations is clearly essential for all future attempts to
experimentally realize the SYK model.    

In this work we propose an extension of the SYK model achieved 
by including a random bilinear term in the Hamiltonian, whose structure
is inspired by experimental considerations \cite{Lantagne2018}. This
additional term leads to a tunable non-Fermi liquid to Fermi liquid 
(nFL/FL) transition similar in some ways to the one seen in
Banerjee-Altman model \cite{Altman2016}. The advantage of our model is
that it is possible to observe such a transition without introducing
an additional flavor of fermions. In addition, the scaling form of the
spectral function can be obtained analytically for the model directly
at the critical point.

 In the following we define our model and discuss the physics that
 drives the nFL/FL  phase transition. We then obtain the saddle point
 equations for the fermion propagator in the limit of large number
 $N$ of fermions and
 analytically extract the low-energy scaling behavior on both sides of the phase
 transition and at the critical point.  We numerically
 solve the saddle point equations and confirm the validity of the
 low-energy scaling forms obtained analytically. 


\section{Model}

\noindent We extend the canonical  SYK model \cite{Kitaev2015} for $N$ Majorana fermions by introducing a random bilinear term as follows
\label{Sec:Setup}
\begin{align}\label{hamiltonian}
H = \sum_{i<j} K_{ij}\chi_i\chi_j + \sum_{i<j<k<l} J_{ijkl} \chi_{i} \chi_{j} \chi_{k} \chi_{l}.
\end{align} 
Here $J_{ijkl}$ denote real coupling constants drawn from a Gaussian random
distribution with $\overline{J_{ijkl}}=0$ and
$\overline{J^2_{ijkl}}=3! J^2/N^3$. The form of the bilinear coupling
$K_{ij}$ is inspired by proposed experimental realizations of the
$\text{SYK}$ model in solid-state systems\cite{Pikulin2017,Alicea2017,Achen2018}  where the randomness in  $J_{ijkl}$ and $K_{ij}$ arises from the disordered spatial structure of the single-particle wavefunctions of the zero modes which comprise the active Majorana degrees of freedom.   Following Ref.\ \onlinecite{Lantagne2018} we take
\begin{align}\label{kij}
K_{ij}=-i\mu \sum_{m=1}^{2M} (a_i^m b_j^m - b_i^m a_j^m)
\end{align}
where $a_i^m,b_j^m$ are random Gaussian independent variables such that $\overline{a_i^m}=\overline{b_i^m}=0$ and $\overline{a_i^ma_j^n} = \overline{b_i^mb_j^n} = \frac{1}{8M}\delta_{ij}\delta^{mn}$ as well as  $\overline{a_i^mb_j^n}=0$ which implies $K^2 = N\overline{K^2_{ij}} = \mu^2/16p $ if we define
\begin{align}\label{p}
 p=M/N.
\end{align}
In the above overline denotes the average over Gaussian distribution.

As noted in Ref.\ \onlinecite{Lantagne2018} model defined by Eq.\
\eqref{hamiltonian} exhibits a phase transition already at the
non-interacting level (i.e.\ when $J=0$). In this case the
single-particle energy spectrum
is given by the eigenvalues of the hermitian matrix $K_{ij}$ defined in
Eq.\ \eqref{kij}. The key observation is that  quantities
$a^m$ and $b^m$ can be viewed as $N$-component vectors in index $i$
and are, for large $N$, approximately orthogonal to each other. As a result, the
rank of the matrix  $K_{ij}$ is close to $4M$. When $4M<N$ the matrix is
rank-deficient and, as a result, has $(N-4M)$ zero modes. One can further
demonstrate\cite{Lantagne2018} that its other eigenvalues cluster
around $\pm\mu/8p$. For $p<p_c={1/4}$ the zero-mode manifold is
separated from the rest of the spectrum by a gap $\Delta$. When interactions
are turned on for $p<p_c$ one might expect that they transform the
degenerate manifold of states at zero energy into an SYK liquid. This
is intuitively clear for weak interactions $J\ll\Delta$ where one
can focus on the zero modes and note that the low energy theory
becomes a canonical SYK model for $(N-4M)$ Majorana fermions in the
zero-mode manifold. As $p$ approaches $p_c$ the gap closes and the
intuitive argument given above fails. In the opposite limit, $p\gg
p_c$, the matrix elements $K_{ij}$ become Gaussian-distributed by
virtue of the central limit theorem. The
non-interacting phase is then a disordered gapless Fermi liquid with a semicircle
spectral density. In this limit it is easy to show that interactions
constitute an irrelevant perturbation. One therefore expects a phase
transition, as a function of increasing parameter $p$, from the SYK liquid to
disordered Fermi liquid in this model.

The model defined by Eqs.\ (\ref{hamiltonian}) and (\ref{kij}) can be
solved, in the large-$N$ limit, by the saddle-point expansion
technique developed for  the original SYK model.\cite{Sachdev2015,Maldacena2016}
Averaging over random variables $a_i^m,b_j^m$  that enter the
definition of $K_{ij}$ involves an extra step that has been described
in Ref.\ \onlinecite{Lantagne2018}.  That study however considered
only the non-interacting model. Here we give the solution for the full
interacting model.  Details fo the calculation are outlined in Appendix A. 
The resulting Matsubara frequency Schwinger-Dyson equation for the fermion propagator $G(\tau)=\frac{1}{N}\sum_i\langle \hat{T} \chi_i(\tau)\chi_i(0) \rangle$  reads
\begin{align}\label{dyson}
G^{-1}(i\omega_n)=-i\omega_n-\Sigma(i\omega_n)
\end{align}
where the self-energy is given by
\begin{align}\label{selfenergy}
\Sigma(i\omega_n) = \Sigma_J(i\omega_n) + 4pK^2\frac{G(i\omega_n)}{4p-K^2G(i\omega_n)^2}.
\end{align} 
The interaction part of the self energy 
\begin{align}\label{selfenergyJ}
\Sigma_J(\tau)=J^2G(\tau)^3
\end{align} 
is the same as in the original Majorana SYK model.

\begin{figure}[t!]
\includegraphics[width=1\columnwidth]{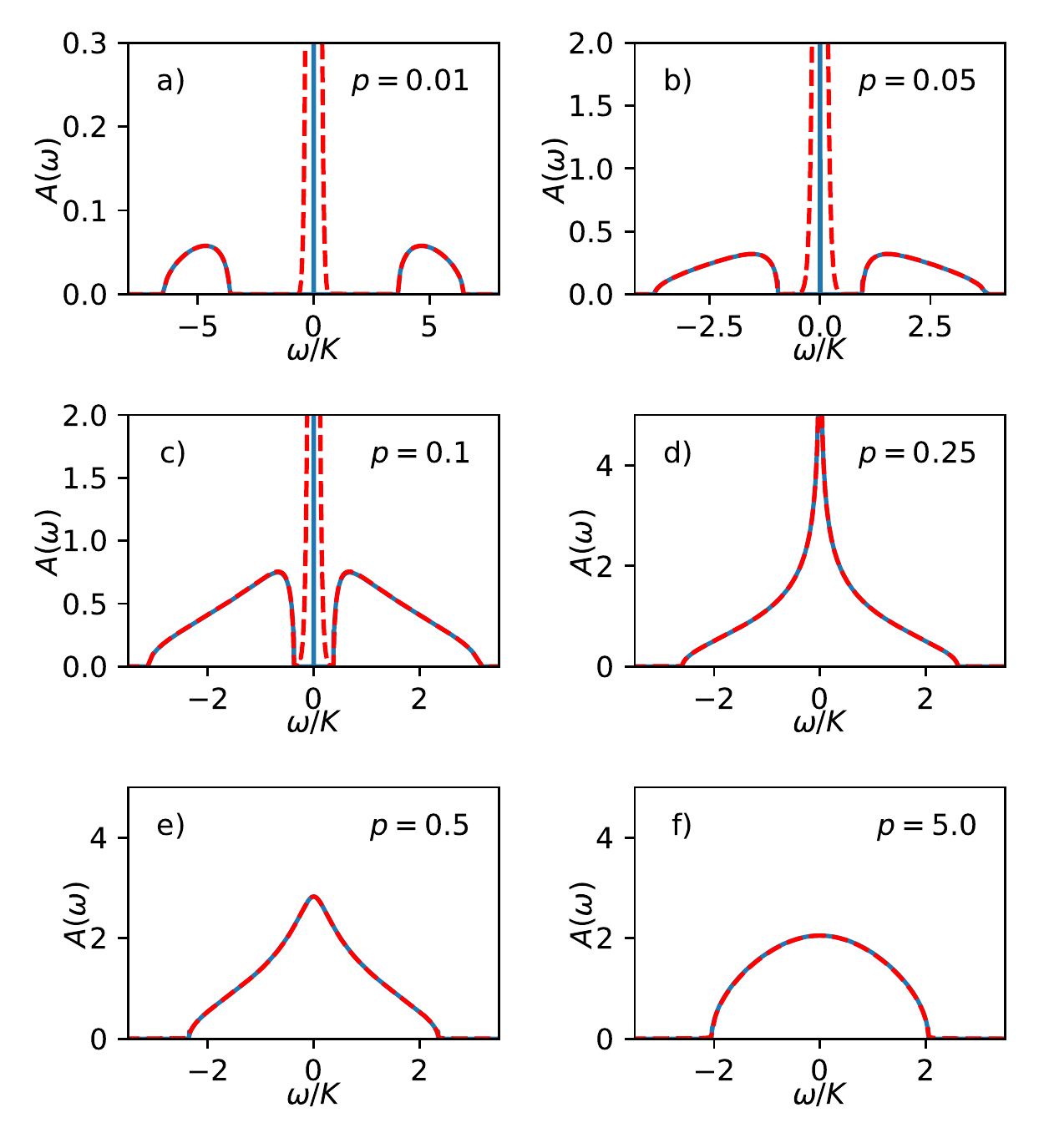}
\caption{Evolution of the spectral function $A(\omega)$ across the
  quantum phase transition tuned by parameter $p=M/N$. Spectral
  functions are computed by a numerical iteration of the large-$N$
  saddle-point equations in Keldysh representation given in Appendix
  B. Results for non-interacting ($J=0$, blue solid line) and
  interacting ($J=K$, red dashed line) cases are overlaid for each
  $p$. Note that the interactions broaden the zero modes
  inside the gap for $p<1/4$, resulting in a SYK non-Fermi
  liquid. Above the transition for $p>1/4$, interactions are
  irrelevant and do not significantly modify the spectral function.}
\label{Fig:linplots}
\end{figure}

\section{Properties of the model}

\subsection{The non-interacting case}
\label{Sec:noninteractingregime}

 We briefly summarize the properties of the non-interacting model. 
The non-interacting problem, obtained by taking  $J=0$ in Eq.\ (\ref{selfenergyJ}), has a simple solution  \cite{Lantagne2018} in the large-$N$ limit. Dropping $\Sigma_J$  equations (\ref{dyson}) and (\ref{selfenergy}) can be combined into a single cubic equation for the propagator in the frequency domain
\begin{align}\label{cubic}
i\omega K^2 G^3 + (1-4p)K^2G^2 - 4ip\omega G - 4p=0.
\end{align}
For $p < 1/4$, the energy spectrum has a gap with a degenerate manifold of zero modes. Contribution of these zero modes to the spectral density is given by $\rho(\omega)=(1-4p)\delta(\omega)$. As $p$ increases, the gap gradually closes until a phase transition occurs at  $p=p_c=1/4$. At the transition the spectral function shows the $A(\omega) \sim \omega^{-1/3}$ scaling. For $p>1/4$ the spectrum remains gapless and approaches the semicircle distribution for large $p$. 

While the Matsubara formalism is useful for analytical calculations we
turn to the Keldysh technique, formulated directly in the real time
domain, to obtain numerical solutions of the large-$N$ saddle point
equations. It allows us to easily extract the spectral function
(without the need to analytically continue from imaginary time) and
the numerical solution converges rapidly which is important when
interactions are included. In the  Keldysh picture propagator  $G$
becomes at $2\times 2$ matrix  with elements $G_{ss'}$ where $s$ represents the contour index for forward and backward paths on the Keldysh contour. Appendix B gives a brief summary of the technique applied to the SYK model while a detailed discussion of the Keldysh formalism can be found in Ref.~\onlinecite{Kamenev2009}.
The Keldysh version of the above cubic equation (\ref{cubic}) reads
\begin{multline}\label{keldyshnoninteract}
    - \omega K^2(\sigma^z G)^3 + K^2(1-4p)(\sigma^z G)^2 \\ +4p\omega \sigma^zG - 4p = 0,
\end{multline}
where $\sigma^z$ acts in the $s$-$s'$ space.
Once the solution is obtained, matrix $G$ contains
$G^T$,$G^<$,$G^{\hat{T}}$ and $G^>$ as its elements. The retarded
Green's function can then be obtained as $G^R(\omega)= G^T(\omega)- G^{<}(\omega)$ which allows us to compute the spectral function $A(\omega)=-2\text{Im}G_R(\omega)$.  Fig.\ \ref{Fig:linplots} shows numerical solutions of Eq.\ (\ref{keldyshnoninteract}) for different values of parameter $p$. These agree with the discussion given below Eq.\ (\ref{cubic}).

\subsection{Interacting regime}
We observed a degenerate manifold of zero modes in the non-interacting regime  for $p<p_c$ separated by a gap from the rest of the spectrum. For weak interaction strength $J$ we can focus on the zero-mode manifold and disregard the rest of the spectrum. The effective low-energy model is then simply an SYK Hamiltonian for the Majorana modes comprising  the zero-mode manifold. We thus expect the $\delta$-function in the spectral density to broaden due to interactions and form a low-energy SYK liquid. Above the non-interacting transition point ($p>p_c$) the single-particle spectrum is gapless. In this case weak four-fermion interactions are known to be irrelevant and we thus expect the interacting system to  form a disordered Fermi liquid in this regime.  

In the two limits  discussed above and at the transition ($p=p_c$) it is possible to analytically extract the low energy scaling behavior of the fermion propagator from Eqs.\ (\ref{dyson}-\ref{selfenergyJ}). In the following, we show that these results indeed confirm the expectations based on the general arguments presented above and we further support these findings by full numerical solutions of the large-$N$ saddle point equations in Keldysh picture.

\subsubsection{Scaling behavior at the transition, $p=p_c=1/4$}
For $p=1/4$, where we observed the transition for the non-interacting system, Eqs.\ \eqref{dyson} and \eqref{selfenergy}  reduce to
\begin{align}\label{e1}
-i\omega - \Sigma_J(i\omega) + \frac{1}{G(i\omega)}\left({1-\frac{1}{K^2G^2(i\omega)}}\right)^{-1} = \frac{1}{G(i\omega)}
\end{align}
In order to obtain the scaling form of $G(i\omega)$, we make a power law ansatz of the form
\begin{align}\label{e2}
 G(i\omega) \sim \omega^{-\alpha}. 
\end{align}
 For $\alpha >0$ we can always go to sufficiently low frequency that $|G| \gg 1/K$ holds.  We can then expand the denominator to first order in $1/(KG)^2$ and obtain
\begin{align}\label{scaling_p0.25}
i\omega \simeq \frac{1}{K^2G^3(i\omega)} -\Sigma_J(i\omega)
\end{align} 
Given the power law ansatz (\ref{e2}), we would like to extract the
scaling form of the SYK self energy defined in  Eq.\
(\ref{selfenergyJ}).  To this end we write the Fourier transform
$G(\tau)=\int d\omega e^{-i\omega \tau}G(i\omega)$. Using a simple result
\begin{align}\label{e22}
\int d\omega e^{-i\omega \tau}\omega^{-\alpha}\sim \tau^{(\alpha-1)}
\end{align}
we find that the
frequency dependence assumed in Eq.\ \eqref{e2} transforms to
$G(\tau) \sim \tau^{(\alpha-1)}$. Eq.\ (\ref{selfenergyJ}) then
implies $\Sigma_J(\tau) \sim \tau^{3(\alpha-1)}$ and the inverse Fourier transform  finally  leads to $\Sigma_J(i\omega)\sim \omega^{2-3\alpha}$. We can now rewrite  Eq.\ \eqref{scaling_p0.25} as
\begin{align}\label{e3}
\omega \simeq A\omega^{3\alpha}  + B\omega^{2-3\alpha}
\end{align}
where we absorbed all prefactors into constants $A$ and $B$. This
equation has a solution for 
$\alpha=1/3$. Therefore, at $p=p_c$ we expect $G\sim
\omega^{-1/3}$.  Remarkably, at the critical point,
the scaling form of the propagator is unchanged by the interactions
even though $\Sigma_J$ enters with the same power in Eq.\ (\ref{e3})
as the non-interacting contribution. Interactions are exactly marginal
at the non-interacting critical point.

\noindent \begin{figure}[t!] 
\includegraphics[width=1\columnwidth]{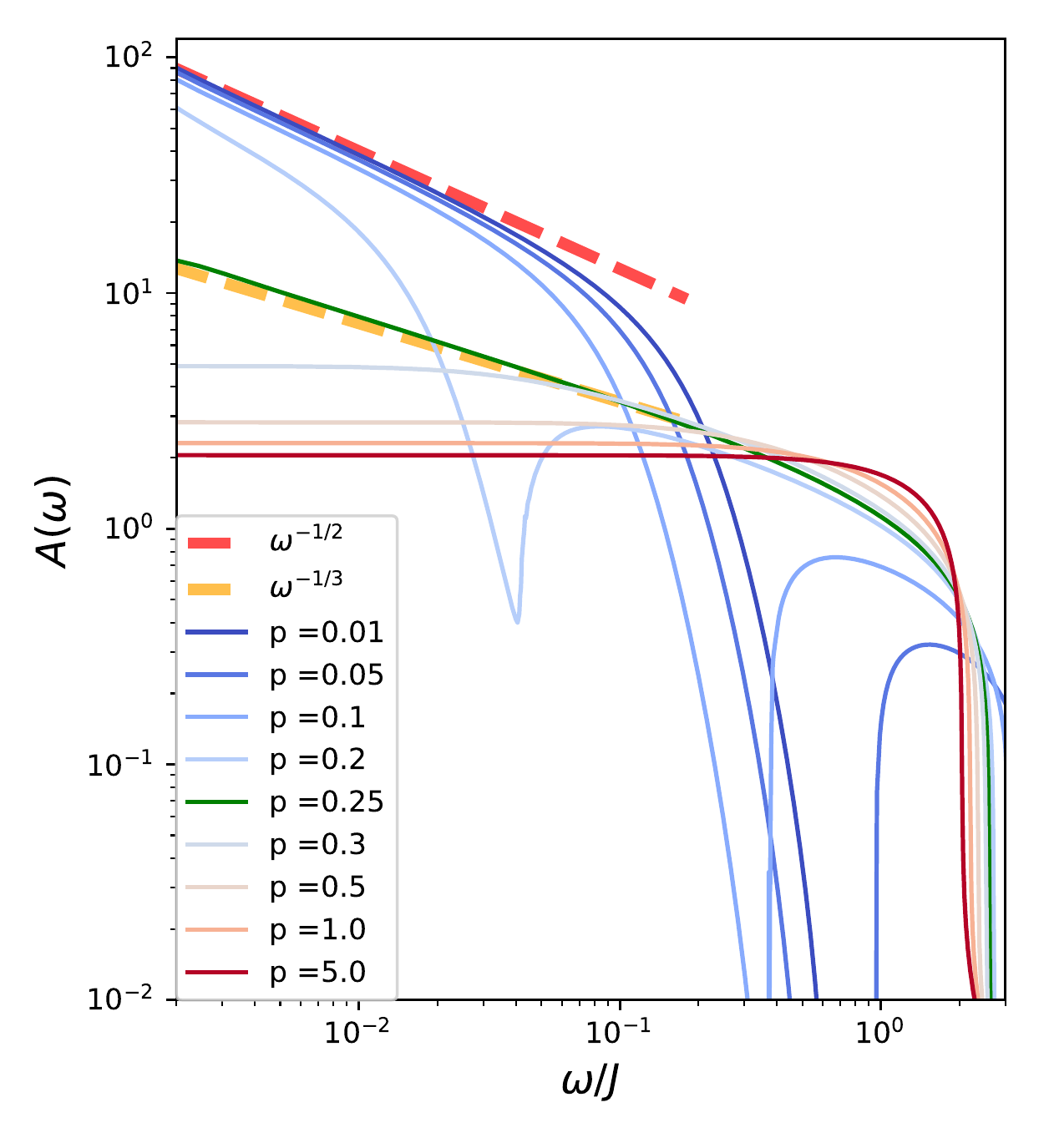}
\caption{Low-frequency scaling behavior of the spectral function
  $A(\omega)$ for various values of $p$. Other parameters $K=J$ and
  $T=0.0003J$ are fixed. Analytically obtained scaling forms are shown
  by dashed lines. For $p<1/4$ spectral function tends to $A(\omega)
  \sim \omega^{-1/2}$, characteristic of the $\text{SYK}$ model. At
  the nFL/FL transition  (green) where $p=1/4$ we confirm that
  $A(\omega) \sim \omega^{-1/3}$ in accord with the prediction. Above the transition
  $A(\omega)$ tends to a frequency-independent constant as expected
  for disordered Fermi liquid.}
\label{Fig:spectralscaling_J}
\end{figure}
\subsubsection{Deep in non-Fermi liquid phase, $p \ll p_c$}
We expect the zero modes to broaden into an SYK peak with $G \sim |\omega|^{-1/2}$ when $p\ll p_c$. In order to see this, we note that in the low-frequency limit we can neglect $p$ in the denominator of Eq.\ \eqref{selfenergy}  which then reduces to 
\begin{align}\label{pll1}
\Sigma(i\omega) \simeq \Sigma_J(i\omega) - 4p\frac{1}{G(i\omega)}.
\end{align}
If we take $J=0$ for a moment, we observe that the expression above
combined with Dyson's equation \eqref{dyson} simplify to
\begin{align}\label{f3}
G(i\omega)=-\frac{1-4p}{i\omega}.
\end{align} 
which gives the spectral density $\rho(\omega) = (1-4p)\delta(\omega)$ after analytic continuation. This is the degenerate manifold of zero modes expected in the absence of interactions. 

Now we turn on the interactions ($J>0$) and investigate how $\Sigma_J$
scales for low frequencies given Eq.\ (\ref{f3}). We find that
$G(\tau) \sim \tau^0$ for $\tau \rightarrow \infty$ which implies that
$\Sigma_J(\tau) \sim \tau^0$ as $\tau \rightarrow \infty$. We can
compare this to the second term $\Sigma_K(i\omega) = -4p/G(i\omega)$
in Eq.\ \eqref{pll1} which is the contribution to the self energy
$\Sigma$ due to the bilinear term in Hamiltonian
\eqref{hamiltonian}. The latter scales as $\Sigma_K(\tau) \sim
\tau^{-2}$, and is therefore dominated by $\Sigma_J \sim \tau^0$ at
long times. We therefore conclude that interactions are relevant and
the manifold of zero modes turns into an SYK liquid where the low
energy solution, characterized by the inverse square root singularity, is well known \cite{Kitaev2015,Sachdev2015}.

\subsubsection{Deep in disordered Fermi liquid phase, $p \gg p_c$}

For large $p \gg p_c$, the self energy in Eq.\ \eqref{selfenergy} reduces to
\begin{align}\label{pgg1} 
\Sigma(i\omega) \simeq \Sigma_J(i\omega) +  K^2G(i\omega)
\end{align} 
If we take $J=0$ and ignore $\Sigma_J$, the exact solution is
available \cite{Pikulin2017} for it becomes a simple quadratic
equation in $G(i\omega)$ and the low energy limit of the solution is
given by $G(i\omega) = i \text{sgn(} \omega \text{)}/K$. This implies
$G(\tau) \sim 1/\tau $ for $ \tau \rightarrow \infty$. As we turn on
interactions, we find $\Sigma_J(\tau) = J^2G^3(\tau) \sim 1/\tau^3 $
which implies  $\Sigma_J(i\omega) \sim \omega^2$, irrelevant at low
frequencies  compared to the second term in Eq.\ \eqref{pgg1}. We
conclude  that for large $p$ the bilinear term dominates and the
spectral function becomes a semicircle characteristic of disordered
Fermi liquid shown in Fig.\ \ref{Fig:linplots}f. Interactions have no
significant effect in this limit.

\subsection{Numerical results}
To confirm the approximate analytical results given above and to extend them beyond the low-frequency regime we solve the Keldysh version of the large-$N$ saddle point equations (4-6) by numerical iteration. The Keldysh saddle point equations are derived in Appendix B and the details of our numerical procedure are described in Appendix C. Fig.\ \ref{Fig:linplots} shows the numerically calculated spectral functions for various values of parameter $p$ over the full range of frequencies while Fig.\ \ref{Fig:spectralscaling_J} focuses on the low-frequency scaling limit. These results are in excellent agreement with analytical scaling forms derived in the preceding subsection and confirm that the phase transition survives the inclusion of strong interactions which transform the gapped phase with degenerate ground state manifold into the SYK non-Fermi liquid.

\section{Summary and conclusions}

We proposed an extension of the Sachdev-Ye-Kitaev model which exhibits
a quantum phase transition from a non-Fermi liquid state to a
disordered Fermi liquid tuned by a dimensionless parameter $p$ which,
in essence, controls the rank of the hermitian matrix in the part of
the Hamiltonian that is bi-linear  in fermion operators. The large-$N$
saddle point equations for the model can be solved analytically in
three limiting cases which establishes the existence of two distinct
stable phases in the model as well as the scale-invariant behavior at the
critical point. These conclusions are confirmed in detail by numerical
solutions of the Keldysh version of the saddle point equations. 

The nFL phase that occurs for $p<p_c$ has an effective description as
a canonical SYK model with $N(1-p/p_c)$ fermions separated by a
gap from the rest of the spectrum. Although we have not computed the
out-of-time order correlator for the model we expect the nFL phase to
saturate the upper bound on the Lyapunov exponent $\lambda\leq 2\pi T$
just like the Banerjee-Altman model\cite{Altman2016}. The $p<p_c$  phase is
therefore expected to be a fast-scrambling, maximally chaotic nFL. Above
the transition we found interactions to be irrelevant and therefore expect this
phase to be slow-scrambling, non-chaotic Fermi liquid. For $p \gg p_c$ case (Eq.\ref{pgg1}) it has been shown \cite{Garcia2018} that $\lambda < 2\pi T$ does not saturate the chaos bound and vanishes below a critical temperature $T^*$.  The model,
therefore, shows a phase transition from fast to slow scrambling
dynamics in what is perhaps the simplest possible setting.

Several extensions of our model are possible and potentially
interesting. The model can be formulated in higher dimensions by
coupling islands described by Hamiltonian \eqref{hamiltonian} via
four-fermion couplings as in Ref.\ \onlinecite{Gu2016}. Such a system
  is then expected to show quantum chaos propagation with a
  characteristic ``butterfly velocity'' in its chaotic phase and exhibit
  a transition to non-chaotic phase as $p$ exceeds the critical
  value. Another obvious extension is to formulate the model with
  complex fermions as in Refs.\ \onlinecite{Sachdev2015,Altman2016}.
  Complex-fermion SYK model exhibits richer behavior because it
  permits the addition of the chemical potential term to control
  fermion density. The phase transition observed in the Majorana
  version of the model studied in this work could show further interesting
  behavior  as a function of density.

\section*{Acknowledgements}
We thank E. Altman, E. Berg, Chengshu Li, \'E. Lantagne-Hurtubise, E. Nica,
A. Nocera and S. Plugge for numerous discussions. The authors
acknowledge support from NSERC and CIfAR. The work reported here was
inspired by conversations held at The Aspen Center for Physics (M.F.)
whose hospitality we would like to acknowledge.
\bibliographystyle{apsrev4-1}
\bibliography{SYK_trans}

\begin{thebibliography}{35}%
\makeatletter
\providecommand \@ifxundefined [1]{%
 \@ifx{#1\undefined}
}%
\providecommand \@ifnum [1]{%
 \ifnum #1\expandafter \@firstoftwo
 \else \expandafter \@secondoftwo
 \fi
}%
\providecommand \@ifx [1]{%
 \ifx #1\expandafter \@firstoftwo
 \else \expandafter \@secondoftwo
 \fi
}%
\providecommand \natexlab [1]{#1}%
\providecommand \enquote  [1]{``#1''}%
\providecommand \bibnamefont  [1]{#1}%
\providecommand \bibfnamefont [1]{#1}%
\providecommand \citenamefont [1]{#1}%
\providecommand \href@noop [0]{\@secondoftwo}%
\providecommand \href [0]{\begingroup \@sanitize@url \@href}%
\providecommand \@href[1]{\@@startlink{#1}\@@href}%
\providecommand \@@href[1]{\endgroup#1\@@endlink}%
\providecommand \@sanitize@url [0]{\catcode `\\12\catcode `\$12\catcode
  `\&12\catcode `\#12\catcode `\^12\catcode `\_12\catcode `\%12\relax}%
\providecommand \@@startlink[1]{}%
\providecommand \@@endlink[0]{}%
\providecommand \url  [0]{\begingroup\@sanitize@url \@url }%
\providecommand \@url [1]{\endgroup\@href {#1}{\urlprefix }}%
\providecommand \urlprefix  [0]{URL }%
\providecommand \Eprint [0]{\href }%
\providecommand \doibase [0]{http://dx.doi.org/}%
\providecommand \selectlanguage [0]{\@gobble}%
\providecommand \bibinfo  [0]{\@secondoftwo}%
\providecommand \bibfield  [0]{\@secondoftwo}%
\providecommand \translation [1]{[#1]}%
\providecommand \BibitemOpen [0]{}%
\providecommand \bibitemStop [0]{}%
\providecommand \bibitemNoStop [0]{.\EOS\space}%
\providecommand \EOS [0]{\spacefactor3000\relax}%
\providecommand \BibitemShut  [1]{\csname bibitem#1\endcsname}%
\let\auto@bib@innerbib\@empty
\bibitem [{\citenamefont {Sachdev}\ and\ \citenamefont {Ye}(1993)}]{SY1996}%
  \BibitemOpen
  \bibfield  {author} {\bibinfo {author} {\bibfnamefont {S.}~\bibnamefont
  {Sachdev}}\ and\ \bibinfo {author} {\bibfnamefont {J.}~\bibnamefont {Ye}},\
  }\href {\doibase 10.1103/PhysRevLett.70.3339} {\bibfield  {journal} {\bibinfo
   {journal} {Phys. Rev. Lett.}\ }\textbf {\bibinfo {volume} {70}},\ \bibinfo
  {pages} {3339} (\bibinfo {year} {1993})}\BibitemShut {NoStop}%
\bibitem [{\citenamefont {Kitaev}(2015)}]{Kitaev2015}%
  \BibitemOpen
  \bibfield  {author} {\bibinfo {author} {\bibfnamefont {A.}~\bibnamefont
  {Kitaev}},\ }\href {http://online.kitp.ucsb.edu/online/entangled15/}
  {\bibfield  {journal} {\bibinfo  {journal} {in KITP Strings Seminar and
  Entanglement 2015 Program}\ } (\bibinfo {year} {2015})}\BibitemShut {NoStop}%
\bibitem [{\citenamefont {Sachdev}(2015)}]{Sachdev2015}%
  \BibitemOpen
  \bibfield  {author} {\bibinfo {author} {\bibfnamefont {S.}~\bibnamefont
  {Sachdev}},\ }\href {\doibase 10.1103/PhysRevX.5.041025} {\bibfield
  {journal} {\bibinfo  {journal} {Phys. Rev. X}\ }\textbf {\bibinfo {volume}
  {5}},\ \bibinfo {pages} {041025} (\bibinfo {year} {2015})}\BibitemShut
  {NoStop}%
\bibitem [{\citenamefont {Maldacena}\ and\ \citenamefont
  {Stanford}(2016)}]{Maldacena2016}%
  \BibitemOpen
  \bibfield  {author} {\bibinfo {author} {\bibfnamefont {J.}~\bibnamefont
  {Maldacena}}\ and\ \bibinfo {author} {\bibfnamefont {D.}~\bibnamefont
  {Stanford}},\ }\href {\doibase 10.1103/PhysRevD.94.106002} {\bibfield
  {journal} {\bibinfo  {journal} {Phys. Rev. D}\ }\textbf {\bibinfo {volume}
  {94}},\ \bibinfo {pages} {106002} (\bibinfo {year} {2016})}\BibitemShut
  {NoStop}%
\bibitem [{\citenamefont {Polchinski}\ and\ \citenamefont
  {Rosenhaus}(2016)}]{Polchinski2016}%
  \BibitemOpen
  \bibfield  {author} {\bibinfo {author} {\bibfnamefont {J.}~\bibnamefont
  {Polchinski}}\ and\ \bibinfo {author} {\bibfnamefont {V.}~\bibnamefont
  {Rosenhaus}},\ }\href {\doibase 10.1007/JHEP04(2016)001} {\bibfield
  {journal} {\bibinfo  {journal} {Journal of High Energy Physics}\ }\textbf
  {\bibinfo {volume} {2016}},\ \bibinfo {pages} {1} (\bibinfo {year}
  {2016})}\BibitemShut {NoStop}%
\bibitem [{\citenamefont {Fu}\ \emph {et~al.}(2017)\citenamefont {Fu},
  \citenamefont {Gaiotto}, \citenamefont {Maldacena},\ and\ \citenamefont
  {Sachdev}}]{Fu2016}%
  \BibitemOpen
  \bibfield  {author} {\bibinfo {author} {\bibfnamefont {W.}~\bibnamefont
  {Fu}}, \bibinfo {author} {\bibfnamefont {D.}~\bibnamefont {Gaiotto}},
  \bibinfo {author} {\bibfnamefont {J.}~\bibnamefont {Maldacena}}, \ and\
  \bibinfo {author} {\bibfnamefont {S.}~\bibnamefont {Sachdev}},\ }\href
  {\doibase 10.1103/PhysRevD.95.026009} {\bibfield  {journal} {\bibinfo
  {journal} {Phys. Rev. D}\ }\textbf {\bibinfo {volume} {95}},\ \bibinfo
  {pages} {026009} (\bibinfo {year} {2017})}\BibitemShut {NoStop}%
\bibitem [{\citenamefont {Witten}(2016)}]{Witten2016}%
  \BibitemOpen
  \bibfield  {author} {\bibinfo {author} {\bibfnamefont {E.}~\bibnamefont
  {Witten}},\ }\href@noop {} {\bibfield  {journal} {\bibinfo  {journal}
  {arXiv:1610.09758}\ } (\bibinfo {year} {2016})}\BibitemShut {NoStop}%
\bibitem [{\citenamefont {Banerjee}\ and\ \citenamefont
  {Altman}(2017)}]{Altman2016}%
  \BibitemOpen
  \bibfield  {author} {\bibinfo {author} {\bibfnamefont {S.}~\bibnamefont
  {Banerjee}}\ and\ \bibinfo {author} {\bibfnamefont {E.}~\bibnamefont
  {Altman}},\ }\href {\doibase 10.1103/PhysRevB.95.134302} {\bibfield
  {journal} {\bibinfo  {journal} {Phys. Rev. B}\ }\textbf {\bibinfo {volume}
  {95}},\ \bibinfo {pages} {134302} (\bibinfo {year} {2017})}\BibitemShut
  {NoStop}%
\bibitem [{\citenamefont {Berkooz}\ \emph {et~al.}(2017)\citenamefont
  {Berkooz}, \citenamefont {Narayan}, \citenamefont {Rozali},\ and\
  \citenamefont {Sim{\'o}n}}]{Berkooz2016}%
  \BibitemOpen
  \bibfield  {author} {\bibinfo {author} {\bibfnamefont {M.}~\bibnamefont
  {Berkooz}}, \bibinfo {author} {\bibfnamefont {P.}~\bibnamefont {Narayan}},
  \bibinfo {author} {\bibfnamefont {M.}~\bibnamefont {Rozali}}, \ and\ \bibinfo
  {author} {\bibfnamefont {J.}~\bibnamefont {Sim{\'o}n}},\ }\href {\doibase
  10.1007/JHEP01(2017)138} {\bibfield  {journal} {\bibinfo  {journal} {Journal
  of High Energy Physics}\ }\textbf {\bibinfo {volume} {2017}},\ \bibinfo
  {pages} {138} (\bibinfo {year} {2017})}\BibitemShut {NoStop}%
\bibitem [{\citenamefont {Bi}\ \emph {et~al.}(2017)\citenamefont {Bi},
  \citenamefont {Jian}, \citenamefont {You}, \citenamefont {Pawlak},\ and\
  \citenamefont {Xu}}]{Bi2017}%
  \BibitemOpen
  \bibfield  {author} {\bibinfo {author} {\bibfnamefont {Z.}~\bibnamefont
  {Bi}}, \bibinfo {author} {\bibfnamefont {C.-M.}\ \bibnamefont {Jian}},
  \bibinfo {author} {\bibfnamefont {Y.-Z.}\ \bibnamefont {You}}, \bibinfo
  {author} {\bibfnamefont {K.~A.}\ \bibnamefont {Pawlak}}, \ and\ \bibinfo
  {author} {\bibfnamefont {C.}~\bibnamefont {Xu}},\ }\href {\doibase
  10.1103/PhysRevB.95.205105} {\bibfield  {journal} {\bibinfo  {journal} {Phys.
  Rev. B}\ }\textbf {\bibinfo {volume} {95}},\ \bibinfo {pages} {205105}
  (\bibinfo {year} {2017})}\BibitemShut {NoStop}%
\bibitem [{\citenamefont {Murugan}\ \emph {et~al.}(2017)\citenamefont
  {Murugan}, \citenamefont {Stanford},\ and\ \citenamefont
  {Witten}}]{Murugan2017}%
  \BibitemOpen
  \bibfield  {author} {\bibinfo {author} {\bibfnamefont {J.}~\bibnamefont
  {Murugan}}, \bibinfo {author} {\bibfnamefont {D.}~\bibnamefont {Stanford}}, \
  and\ \bibinfo {author} {\bibfnamefont {E.}~\bibnamefont {Witten}},\ }\href
  {\doibase 10.1007/JHEP08(2017)146} {\bibfield  {journal} {\bibinfo  {journal}
  {J. High Energy Phys.}\ }\textbf {\bibinfo {volume} {2017}},\ \bibinfo
  {pages} {146} (\bibinfo {year} {2017})}\BibitemShut {NoStop}%
\bibitem [{\citenamefont {Peng}\ \emph {et~al.}(2017)\citenamefont {Peng},
  \citenamefont {Spradlin},\ and\ \citenamefont {Volovich}}]{Peng2017}%
  \BibitemOpen
  \bibfield  {author} {\bibinfo {author} {\bibfnamefont {C.}~\bibnamefont
  {Peng}}, \bibinfo {author} {\bibfnamefont {M.}~\bibnamefont {Spradlin}}, \
  and\ \bibinfo {author} {\bibfnamefont {A.}~\bibnamefont {Volovich}},\ }\href
  {\doibase 10.1007/JHEP05(2017)062} {\bibfield  {journal} {\bibinfo  {journal}
  {J. High Energy Phys.}\ }\textbf {\bibinfo {volume} {2017}},\ \bibinfo
  {pages} {62} (\bibinfo {year} {2017})}\BibitemShut {NoStop}%
\bibitem [{\citenamefont {Lantagne-Hurtubise}\ \emph
  {et~al.}(2018)\citenamefont {Lantagne-Hurtubise}, \citenamefont {Li},\ and\
  \citenamefont {Franz}}]{Lantagne2018}%
  \BibitemOpen
  \bibfield  {author} {\bibinfo {author} {\bibfnamefont {E.}~\bibnamefont
  {Lantagne-Hurtubise}}, \bibinfo {author} {\bibfnamefont {C.}~\bibnamefont
  {Li}}, \ and\ \bibinfo {author} {\bibfnamefont {M.}~\bibnamefont {Franz}},\
  }\href {\doibase 10.1103/PhysRevB.97.235124} {\bibfield  {journal} {\bibinfo
  {journal} {Phys. Rev. B}\ }\textbf {\bibinfo {volume} {97}},\ \bibinfo
  {pages} {235124} (\bibinfo {year} {2018})}\BibitemShut {NoStop}%
\bibitem [{\citenamefont {Liu}\ \emph {et~al.}(2018)\citenamefont {Liu},
  \citenamefont {Chen},\ and\ \citenamefont {Balents}}]{Liu2017}%
  \BibitemOpen
  \bibfield  {author} {\bibinfo {author} {\bibfnamefont {C.}~\bibnamefont
  {Liu}}, \bibinfo {author} {\bibfnamefont {X.}~\bibnamefont {Chen}}, \ and\
  \bibinfo {author} {\bibfnamefont {L.}~\bibnamefont {Balents}},\ }\href
  {\doibase 10.1103/PhysRevB.97.245126} {\bibfield  {journal} {\bibinfo
  {journal} {Phys. Rev. B}\ }\textbf {\bibinfo {volume} {97}},\ \bibinfo
  {pages} {245126} (\bibinfo {year} {2018})}\BibitemShut {NoStop}%
\bibitem [{\citenamefont {Song}\ \emph {et~al.}(2017)\citenamefont {Song},
  \citenamefont {Jian},\ and\ \citenamefont {Balents}}]{Balents2017}%
  \BibitemOpen
  \bibfield  {author} {\bibinfo {author} {\bibfnamefont {X.-Y.}\ \bibnamefont
  {Song}}, \bibinfo {author} {\bibfnamefont {C.-M.}\ \bibnamefont {Jian}}, \
  and\ \bibinfo {author} {\bibfnamefont {L.}~\bibnamefont {Balents}},\ }\href
  {\doibase 10.1103/PhysRevLett.119.216601} {\bibfield  {journal} {\bibinfo
  {journal} {Phys. Rev. Lett.}\ }\textbf {\bibinfo {volume} {119}},\ \bibinfo
  {pages} {216601} (\bibinfo {year} {2017})}\BibitemShut {NoStop}%
\bibitem [{\citenamefont {{Huang}}\ and\ \citenamefont
  {{Gu}}(2017)}]{Huang2017}%
  \BibitemOpen
  \bibfield  {author} {\bibinfo {author} {\bibfnamefont {Y.}~\bibnamefont
  {{Huang}}}\ and\ \bibinfo {author} {\bibfnamefont {Y.}~\bibnamefont {{Gu}}},\
  }\href@noop {} {\bibfield  {journal} {\bibinfo  {journal} {ArXiv e-prints}\ }
  (\bibinfo {year} {2017})},\ \Eprint {http://arxiv.org/abs/1709.09160}
  {arXiv:1709.09160 [hep-th]} \BibitemShut {NoStop}%
\bibitem [{\citenamefont {Jian}\ \emph {et~al.}(2017)\citenamefont {Jian},
  \citenamefont {Bi},\ and\ \citenamefont {Xu}}]{CenkeXu2017a}%
  \BibitemOpen
  \bibfield  {author} {\bibinfo {author} {\bibfnamefont {C.-M.}\ \bibnamefont
  {Jian}}, \bibinfo {author} {\bibfnamefont {Z.}~\bibnamefont {Bi}}, \ and\
  \bibinfo {author} {\bibfnamefont {C.}~\bibnamefont {Xu}},\ }\href {\doibase
  10.1103/PhysRevB.96.115122} {\bibfield  {journal} {\bibinfo  {journal} {Phys.
  Rev. B}\ }\textbf {\bibinfo {volume} {96}},\ \bibinfo {pages} {115122}
  (\bibinfo {year} {2017})}\BibitemShut {NoStop}%
\bibitem [{\citenamefont {Wu}\ \emph {et~al.}(2018)\citenamefont {Wu},
  \citenamefont {Chen}, \citenamefont {Jian}, \citenamefont {You},\ and\
  \citenamefont {Xu}}]{CenkeXu2018a}%
  \BibitemOpen
  \bibfield  {author} {\bibinfo {author} {\bibfnamefont {X.}~\bibnamefont
  {Wu}}, \bibinfo {author} {\bibfnamefont {X.}~\bibnamefont {Chen}}, \bibinfo
  {author} {\bibfnamefont {C.-M.}\ \bibnamefont {Jian}}, \bibinfo {author}
  {\bibfnamefont {Y.-Z.}\ \bibnamefont {You}}, \ and\ \bibinfo {author}
  {\bibfnamefont {C.}~\bibnamefont {Xu}},\ }\href {\doibase
  10.1103/PhysRevB.98.165117} {\bibfield  {journal} {\bibinfo  {journal} {Phys.
  Rev. B}\ }\textbf {\bibinfo {volume} {98}},\ \bibinfo {pages} {165117}
  (\bibinfo {year} {2018})}\BibitemShut {NoStop}%
\bibitem [{\citenamefont {Hosur}\ \emph {et~al.}(2016)\citenamefont {Hosur},
  \citenamefont {Qi}, \citenamefont {Roberts},\ and\ \citenamefont
  {Yoshida}}]{Hosur2016}%
  \BibitemOpen
  \bibfield  {author} {\bibinfo {author} {\bibfnamefont {P.}~\bibnamefont
  {Hosur}}, \bibinfo {author} {\bibfnamefont {X.-L.}\ \bibnamefont {Qi}},
  \bibinfo {author} {\bibfnamefont {D.~A.}\ \bibnamefont {Roberts}}, \ and\
  \bibinfo {author} {\bibfnamefont {B.}~\bibnamefont {Yoshida}},\ }\href
  {\doibase 10.1007/JHEP02(2016)004} {\bibfield  {journal} {\bibinfo  {journal}
  {Journal of High Energy Physics}\ }\textbf {\bibinfo {volume} {2016}},\
  \bibinfo {pages} {4} (\bibinfo {year} {2016})}\BibitemShut {NoStop}%
\bibitem [{\citenamefont {Gu}\ \emph {et~al.}(2017)\citenamefont {Gu},
  \citenamefont {Qi},\ and\ \citenamefont {Stanford}}]{Gu2016}%
  \BibitemOpen
  \bibfield  {author} {\bibinfo {author} {\bibfnamefont {Y.}~\bibnamefont
  {Gu}}, \bibinfo {author} {\bibfnamefont {X.-L.}\ \bibnamefont {Qi}}, \ and\
  \bibinfo {author} {\bibfnamefont {D.}~\bibnamefont {Stanford}},\ }\href
  {\doibase 10.1007/JHEP05(2017)125} {\bibfield  {journal} {\bibinfo  {journal}
  {Journal of High Energy Physics}\ }\textbf {\bibinfo {volume} {2017}},\
  \bibinfo {pages} {125} (\bibinfo {year} {2017})}\BibitemShut {NoStop}%
\bibitem [{\citenamefont {Chen}\ \emph {et~al.}(2017)\citenamefont {Chen},
  \citenamefont {Zhai},\ and\ \citenamefont {Zhang}}]{Chen2017}%
  \BibitemOpen
  \bibfield  {author} {\bibinfo {author} {\bibfnamefont {Y.}~\bibnamefont
  {Chen}}, \bibinfo {author} {\bibfnamefont {H.}~\bibnamefont {Zhai}}, \ and\
  \bibinfo {author} {\bibfnamefont {P.}~\bibnamefont {Zhang}},\ }\href
  {\doibase 10.1007/JHEP07(2017)150} {\bibfield  {journal} {\bibinfo  {journal}
  {J. High Energy Phys.}\ }\textbf {\bibinfo {volume} {2017}},\ \bibinfo
  {pages} {150} (\bibinfo {year} {2017})}\BibitemShut {NoStop}%
\bibitem [{\citenamefont {Krishnan}\ \emph {et~al.}(2017)\citenamefont
  {Krishnan}, \citenamefont {Sanyal},\ and\ \citenamefont
  {Subramanian}}]{Krishnan2017}%
  \BibitemOpen
  \bibfield  {author} {\bibinfo {author} {\bibfnamefont {C.}~\bibnamefont
  {Krishnan}}, \bibinfo {author} {\bibfnamefont {S.}~\bibnamefont {Sanyal}}, \
  and\ \bibinfo {author} {\bibfnamefont {P.~N.~B.}\ \bibnamefont
  {Subramanian}},\ }\href {\doibase 10.1007/JHEP03(2017)056} {\bibfield
  {journal} {\bibinfo  {journal} {J. High Energy Phys.}\ }\textbf {\bibinfo
  {volume} {2017}},\ \bibinfo {pages} {56} (\bibinfo {year}
  {2017})}\BibitemShut {NoStop}%
\bibitem [{\citenamefont {Garc\'{\i}a-\'Alvarez}\ \emph
  {et~al.}(2017)\citenamefont {Garc\'{\i}a-\'Alvarez}, \citenamefont
  {Egusquiza}, \citenamefont {Lamata}, \citenamefont {del Campo}, \citenamefont
  {Sonner},\ and\ \citenamefont {Solano}}]{Solano2017}%
  \BibitemOpen
  \bibfield  {author} {\bibinfo {author} {\bibfnamefont {L.}~\bibnamefont
  {Garc\'{\i}a-\'Alvarez}}, \bibinfo {author} {\bibfnamefont {I.~L.}\
  \bibnamefont {Egusquiza}}, \bibinfo {author} {\bibfnamefont {L.}~\bibnamefont
  {Lamata}}, \bibinfo {author} {\bibfnamefont {A.}~\bibnamefont {del Campo}},
  \bibinfo {author} {\bibfnamefont {J.}~\bibnamefont {Sonner}}, \ and\ \bibinfo
  {author} {\bibfnamefont {E.}~\bibnamefont {Solano}},\ }\href {\doibase
  10.1103/PhysRevLett.119.040501} {\bibfield  {journal} {\bibinfo  {journal}
  {Phys. Rev. Lett.}\ }\textbf {\bibinfo {volume} {119}},\ \bibinfo {pages}
  {040501} (\bibinfo {year} {2017})}\BibitemShut {NoStop}%
\bibitem [{\citenamefont {Luo}\ \emph {et~al.}(2017)\citenamefont {Luo},
  \citenamefont {You}, \citenamefont {Li}, \citenamefont {Jian}, \citenamefont
  {Lu}, \citenamefont {Xu}, \citenamefont {Zeng},\ and\ \citenamefont
  {Laflamme}}]{Laflamme2017}%
  \BibitemOpen
  \bibfield  {author} {\bibinfo {author} {\bibfnamefont {Z.}~\bibnamefont
  {Luo}}, \bibinfo {author} {\bibfnamefont {Y.-Z.}\ \bibnamefont {You}},
  \bibinfo {author} {\bibfnamefont {J.}~\bibnamefont {Li}}, \bibinfo {author}
  {\bibfnamefont {C.-M.}\ \bibnamefont {Jian}}, \bibinfo {author}
  {\bibfnamefont {D.}~\bibnamefont {Lu}}, \bibinfo {author} {\bibfnamefont
  {C.}~\bibnamefont {Xu}}, \bibinfo {author} {\bibfnamefont {B.}~\bibnamefont
  {Zeng}}, \ and\ \bibinfo {author} {\bibfnamefont {R.}~\bibnamefont
  {Laflamme}},\ }\href@noop {} {\  (\bibinfo {year} {2017})},\ \Eprint
  {http://arxiv.org/abs/1712.06458} {arXiv:1712.06458 [quant-ph]} \BibitemShut
  {NoStop}%
\bibitem [{\citenamefont {You}\ \emph {et~al.}(2017)\citenamefont {You},
  \citenamefont {Ludwig},\ and\ \citenamefont {Xu}}]{Xu2016}%
  \BibitemOpen
  \bibfield  {author} {\bibinfo {author} {\bibfnamefont {Y.-Z.}\ \bibnamefont
  {You}}, \bibinfo {author} {\bibfnamefont {A.~W.~W.}\ \bibnamefont {Ludwig}},
  \ and\ \bibinfo {author} {\bibfnamefont {C.}~\bibnamefont {Xu}},\ }\href
  {\doibase 10.1103/PhysRevB.95.115150} {\bibfield  {journal} {\bibinfo
  {journal} {Phys. Rev. B}\ }\textbf {\bibinfo {volume} {95}},\ \bibinfo
  {pages} {115150} (\bibinfo {year} {2017})}\BibitemShut {NoStop}%
\bibitem [{\citenamefont {Garc\'{\i}a-Garc\'{\i}a}\ and\ \citenamefont
  {Verbaarschot}(2016)}]{Verbaar2016}%
  \BibitemOpen
  \bibfield  {author} {\bibinfo {author} {\bibfnamefont {A.~M.}\ \bibnamefont
  {Garc\'{\i}a-Garc\'{\i}a}}\ and\ \bibinfo {author} {\bibfnamefont {J.~J.~M.}\
  \bibnamefont {Verbaarschot}},\ }\href {\doibase 10.1103/PhysRevD.94.126010}
  {\bibfield  {journal} {\bibinfo  {journal} {Phys. Rev. D}\ }\textbf {\bibinfo
  {volume} {94}},\ \bibinfo {pages} {126010} (\bibinfo {year}
  {2016})}\BibitemShut {NoStop}%
\bibitem [{\citenamefont {Li}\ \emph {et~al.}(2017)\citenamefont {Li},
  \citenamefont {Liu}, \citenamefont {Xin},\ and\ \citenamefont
  {Zhou}}]{Li2017}%
  \BibitemOpen
  \bibfield  {author} {\bibinfo {author} {\bibfnamefont {T.}~\bibnamefont
  {Li}}, \bibinfo {author} {\bibfnamefont {J.}~\bibnamefont {Liu}}, \bibinfo
  {author} {\bibfnamefont {Y.}~\bibnamefont {Xin}}, \ and\ \bibinfo {author}
  {\bibfnamefont {Y.}~\bibnamefont {Zhou}},\ }\href {\doibase
  10.1007/JHEP06(2017)111} {\bibfield  {journal} {\bibinfo  {journal} {J. High
  Energy Phys.}\ }\textbf {\bibinfo {volume} {2017}},\ \bibinfo {pages} {111}
  (\bibinfo {year} {2017})}\BibitemShut {NoStop}%
\bibitem [{\citenamefont {Jian}\ and\ \citenamefont {Yao}(2017)}]{Jian2017}%
  \BibitemOpen
  \bibfield  {author} {\bibinfo {author} {\bibfnamefont {S.-K.}\ \bibnamefont
  {Jian}}\ and\ \bibinfo {author} {\bibfnamefont {H.}~\bibnamefont {Yao}},\
  }\href {\doibase 10.1103/PhysRevLett.119.206602} {\bibfield  {journal}
  {\bibinfo  {journal} {Phys. Rev. Lett.}\ }\textbf {\bibinfo {volume} {119}},\
  \bibinfo {pages} {206602} (\bibinfo {year} {2017})}\BibitemShut {NoStop}%
\bibitem [{\citenamefont {Maldacena}\ and\ \citenamefont
  {Qi}(2018)}]{Maldacena:2018}%
  \BibitemOpen
  \bibfield  {author} {\bibinfo {author} {\bibfnamefont {J.}~\bibnamefont
  {Maldacena}}\ and\ \bibinfo {author} {\bibfnamefont {X.-L.}\ \bibnamefont
  {Qi}},\ }\href@noop {} {\  (\bibinfo {year} {2018})},\ \Eprint
  {http://arxiv.org/abs/1804.00491} {arXiv:1804.00491 [hep-th]} \BibitemShut
  {NoStop}%
\bibitem [{\citenamefont {Danshita}\ \emph {et~al.}(2017)\citenamefont
  {Danshita}, \citenamefont {Hanada},\ and\ \citenamefont
  {Tezuka}}]{Danshita2017}%
  \BibitemOpen
  \bibfield  {author} {\bibinfo {author} {\bibfnamefont {I.}~\bibnamefont
  {Danshita}}, \bibinfo {author} {\bibfnamefont {M.}~\bibnamefont {Hanada}}, \
  and\ \bibinfo {author} {\bibfnamefont {M.}~\bibnamefont {Tezuka}},\ }\href
  {\doibase 10.1093/ptep/ptx108} {\bibfield  {journal} {\bibinfo  {journal}
  {Progr. Theor. Exp. Phys.}\ }\textbf {\bibinfo {volume} {2017}},\ \bibinfo
  {pages} {083I01} (\bibinfo {year} {2017})}\BibitemShut {NoStop}%
\bibitem [{\citenamefont {Pikulin}\ and\ \citenamefont
  {Franz}(2017)}]{Pikulin2017}%
  \BibitemOpen
  \bibfield  {author} {\bibinfo {author} {\bibfnamefont {D.~I.}\ \bibnamefont
  {Pikulin}}\ and\ \bibinfo {author} {\bibfnamefont {M.}~\bibnamefont
  {Franz}},\ }\href {\doibase 10.1103/PhysRevX.7.031006} {\bibfield  {journal}
  {\bibinfo  {journal} {Phys. Rev. X}\ }\textbf {\bibinfo {volume} {7}},\
  \bibinfo {pages} {031006} (\bibinfo {year} {2017})}\BibitemShut {NoStop}%
\bibitem [{\citenamefont {Garc\'{\i}a-Garc\'{\i}a}\ \emph
  {et~al.}(2018)\citenamefont {Garc\'{\i}a-Garc\'{\i}a}, \citenamefont
  {Loureiro}, \citenamefont {Romero-Berm\'udez},\ and\ \citenamefont
  {Tezuka}}]{Garcia2018}%
  \BibitemOpen
  \bibfield  {author} {\bibinfo {author} {\bibfnamefont {A.~M.}\ \bibnamefont
  {Garc\'{\i}a-Garc\'{\i}a}}, \bibinfo {author} {\bibfnamefont
  {B.}~\bibnamefont {Loureiro}}, \bibinfo {author} {\bibfnamefont
  {A.}~\bibnamefont {Romero-Berm\'udez}}, \ and\ \bibinfo {author}
  {\bibfnamefont {M.}~\bibnamefont {Tezuka}},\ }\href {\doibase
  10.1103/PhysRevLett.120.241603} {\bibfield  {journal} {\bibinfo  {journal}
  {Phys. Rev. Lett.}\ }\textbf {\bibinfo {volume} {120}},\ \bibinfo {pages}
  {241603} (\bibinfo {year} {2018})}\BibitemShut {NoStop}%
\bibitem [{\citenamefont {Chew}\ \emph {et~al.}(2017)\citenamefont {Chew},
  \citenamefont {Essin},\ and\ \citenamefont {Alicea}}]{Alicea2017}%
  \BibitemOpen
  \bibfield  {author} {\bibinfo {author} {\bibfnamefont {A.}~\bibnamefont
  {Chew}}, \bibinfo {author} {\bibfnamefont {A.}~\bibnamefont {Essin}}, \ and\
  \bibinfo {author} {\bibfnamefont {J.}~\bibnamefont {Alicea}},\ }\href
  {\doibase 10.1103/PhysRevB.96.121119} {\bibfield  {journal} {\bibinfo
  {journal} {Phys. Rev. B}\ }\textbf {\bibinfo {volume} {96}},\ \bibinfo
  {pages} {121119} (\bibinfo {year} {2017})}\BibitemShut {NoStop}%
\bibitem [{\citenamefont {Chen}\ \emph {et~al.}(2018)\citenamefont {Chen},
  \citenamefont {Ilan}, \citenamefont {de~Juan}, \citenamefont {Pikulin},\ and\
  \citenamefont {Franz}}]{Achen2018}%
  \BibitemOpen
  \bibfield  {author} {\bibinfo {author} {\bibfnamefont {A.}~\bibnamefont
  {Chen}}, \bibinfo {author} {\bibfnamefont {R.}~\bibnamefont {Ilan}}, \bibinfo
  {author} {\bibfnamefont {F.}~\bibnamefont {de~Juan}}, \bibinfo {author}
  {\bibfnamefont {D.~I.}\ \bibnamefont {Pikulin}}, \ and\ \bibinfo {author}
  {\bibfnamefont {M.}~\bibnamefont {Franz}},\ }\href {\doibase
  10.1103/PhysRevLett.121.036403} {\bibfield  {journal} {\bibinfo  {journal}
  {Phys. Rev. Lett.}\ }\textbf {\bibinfo {volume} {121}},\ \bibinfo {pages}
  {036403} (\bibinfo {year} {2018})}\BibitemShut {NoStop}%
\bibitem [{\citenamefont {Kamenev}\ and\ \citenamefont
  {Levchenko}(2009)}]{Kamenev2009}%
  \BibitemOpen
  \bibfield  {author} {\bibinfo {author} {\bibfnamefont {A.}~\bibnamefont
  {Kamenev}}\ and\ \bibinfo {author} {\bibfnamefont {A.}~\bibnamefont
  {Levchenko}},\ }\href {\doibase 10.1080/00018730902850504} {\bibfield
  {journal} {\bibinfo  {journal} {Advances in Physics}\ }\textbf {\bibinfo
  {volume} {58}},\ \bibinfo {pages} {197} (\bibinfo {year} {2009})},\ \Eprint
  {http://arxiv.org/abs/https://doi.org/10.1080/00018730902850504}
  {https://doi.org/10.1080/00018730902850504} \BibitemShut {NoStop}%
\bibitem [{\citenamefont {Jones}\ \emph {et~al.}(01  )\citenamefont {Jones},
  \citenamefont {Oliphant}, \citenamefont {Peterson} \emph {et~al.}}]{scipy}%
  \BibitemOpen
  \bibfield  {author} {\bibinfo {author} {\bibfnamefont {E.}~\bibnamefont
  {Jones}}, \bibinfo {author} {\bibfnamefont {T.}~\bibnamefont {Oliphant}},
  \bibinfo {author} {\bibfnamefont {P.}~\bibnamefont {Peterson}},  \emph
  {et~al.},\ }\href {http://www.scipy.org/} {\enquote {\bibinfo {title}
  {{SciPy}: Open source scientific tools for {Python}},}\ } (\bibinfo {year}
  {2001--}),\ \bibinfo {note} {}\BibitemShut
  {NoStop}%
\end{thebibliography}%
\begin{widetext}
\section*{APPENDIX}
\subsection{Large-$N$ solution via imaginary-time path integral}

We follow the standard procedure outlined in Ref.\ \onlinecite{Maldacena2016}, the so called
$G$-$\Sigma$ formalism, which leads to saddle point equations for the
fermion propagator and the self energy that become asymptotically
exact in the limit of large number of fermions $N$. As the first step
we reformulate the problem defined by Hamiltonian (\ref{hamiltonian}) as an imaginary time path
integral for the partition function  $Z=\int \mathcal{D}[\chi] e^{-S}$
where the action reads
\begin{align}
S [\chi]=\int d\tau \left[\frac{1}{2}\sum_i\chi_i \partial \chi_i-i\mu\sum_{ijm}a_i^mb_j^m\chi_i\chi_j + \sum_{ijkl}J_{ijkl}\chi_i\chi_j\chi_k\chi_l \right].
\end{align}
To decouple the random variables $a^m_i$ and $b^m_j$ we  employ the
identity $\int \mathcal{D} [\psi_a, \psi_b] \exp\left(-i\mu\int d\tau
  \sum_m \psi^m_a \psi^m_b \right) = 1$ where $\psi^m_a$ and  $\psi^m_b$
are auxiliary Grassman variables. After change of variables
$\psi_a^m \rightarrow \psi_a^m + \sum_i a_i^m \chi_i$ and $\psi_b^m
\rightarrow \psi_b^m + \sum_j b_j^m \chi_j$  we obtain 
\begin{align}\label{a5}
S[\psi_a, \psi_b, \chi] = \int d\tau \left[ \frac{1}{2}
  \chi_i \partial_\tau \chi_i + i \mu \left(  \psi_a^m \psi_b^m -
  a_i^m \psi_b^m \chi_i  + b_j^m \psi_a^m \chi_j \right) +  \sum_{ijkl}J_{ijkl}\chi_i\chi_j\chi_k\chi_l  \right].
\end{align}
It is now straightforward to  perform the Gaussian average over the random variables which leads
to an action that is bi-local in time variable
\begin{multline*}
S= \int d\tau \left[ \frac{1}{2}\sum_i\chi_i \partial_\tau \chi_i
+i\mu\sum_m
\psi^m_a \psi^m_b \right] -
\frac{K^2}{N} \int d\tau d\tau' \left[\sum_m \psi_b^m(\tau)\psi_b^m(\tau')\sum_i \chi_i(\tau)\chi_i(\tau') \right] + (a \leftrightarrow b)  \\
  -  \frac{J^2}{8N^3}\int d\tau d\tau' \left(\sum_i\chi_i(\tau)\chi_i(\tau')\right)^4.
\end{multline*}
More details on the steps above can be found in  Ref.~\onlinecite{Lantagne2018}.
We next introduce propagators for fermionic degrees of freedom
using  bosonic path integral identities
$
\int \mathcal{D}[\Sigma]\exp\left(-\frac{N}{2}\Sigma(\tau,\tau')\left[
G(\tau,\tau')-\frac{1}{N}\sum_i\chi_i(\tau)\chi(\tau')\right]\right) = 1
$
and 
$
\int \mathcal{D}[\Omega]\exp\left(-M\Omega^{\alpha\beta}(\tau,\tau')\left[
F^{ \beta \alpha}(\tau,\tau')-\frac{1}{2M}\sum_m\psi^m_\alpha(\tau)\psi_\beta^m(\tau')\right]\right) = 1
$ 
where $\Sigma$ and $\Omega$  act as Lagrange multipliers and have
physical interpretation as fermion self energies. After integrating
out fermions we obtain the saddle-point action \begin{equation}\label{act22}
S= \frac{N}{2}\text{tr}\log \left[-\partial_\tau + \Sigma \right] + M
\text{tr} \log \left[ \mu\sigma^y + \Omega \right] 
 +\int d\tau d\tau'\left[ \frac{N}{2}\Sigma G  +M\Omega^{\alpha\beta}F^{ \beta \alpha}
+2K^2M (F^{aa}+F^{bb})G  -  \frac{J^2}{8}N G^4
    \right]
\end{equation} 
\end{widetext}
where we suppressed temporal dependence for the sake of
brevity. Indices $\alpha, \beta$ take values $a, b$ and summation over
repeated indices is assumed. We
note that except for the last term, which incorporates the effect of
interactions,  the action (\ref{act22}) has the same form as the action derived in
Appendix A of Ref.\ \onlinecite{Lantagne2018}.

Varying the action with respect to the self energies gives the two Dyson
equations,
\begin{gather}
G(i\omega_n)(i\omega_n-\Sigma(i\omega_n))=1 \label{matsp4} 
\\
\sum_\gamma F^{\alpha \gamma}(i\omega)\left(\mu\sigma^y_{\gamma \beta}
  + \Omega^{\gamma \beta}(i \omega) \right) = -\delta^{\alpha \beta}.
 \label{matsp1}
\end{gather}
Varying with respect to the propagators yields the saddle-point equations
for self energies,
\begin{gather}
\Sigma(\tau, \tau') =  J^2G(\tau,\tau')^3 - 4pK^2F^{\gamma \gamma}(\tau, \tau')  \label{matsp2} \\
\Omega ^{\alpha\beta} (i\omega) = -2K^2\delta^{\alpha\beta}G(i\omega) \label{matsp3}.
\end{gather}
Notice that $\Omega^{\alpha \beta}=\Omega\delta_{\alpha\beta}$ is diagonal according to Eq.\
(\ref{matsp3}) where we defined $\Omega=-2K^2G$. This ensures that $F^{aa} = F^{bb}\equiv F$ which can be seen by explicitly writing \eqref{matsp1} as a matrix product. Then equation \eqref{matsp1} can be simplified to
\begin{equation}
F(\mu^2-\Omega^2)=\Omega
\end{equation}
which can be seen by explicitly writing out the matrix product. Substituting $\Omega=-2K^2G$ we then obtain
\begin{equation}
F = -\frac{G}{2(4p-K^2G^2)}.
\end{equation} 
Finally substituting this into Eq.\ \eqref{matsp2} (notice the repeated indices) we obtain the self-energy expression \eqref{selfenergy}
to be solved in combination with equation \eqref{matsp4}. Note that from \eqref{matsp2} we defined $\Sigma_J(\tau, \tau') =  J^2G(\tau,\tau')^3$.
\\
\subsection{Keldysh Path Integral}
To derive the Keldysh version of large-$N$ saddle point equations  we
follow Ref.~\onlinecite{Balents2017}. (For an introduction to Keldysh formalism see e.g.\ Ref.~\onlinecite{Kamenev2009}). We write down the Keldysh path
integral for Hamiltonian \eqref{hamiltonian} and disorder-average the
partition fuction $Z$ over Gaussian random variables of the
model. This process is almost identical to the disorder averaging of
SYK models in Matsubara formalism \cite{Maldacena2016}. The
non-standard form of the bilinear coupling $K_{ij}$ is handled in a
fashion nearly identical to Appendix A above. Disorder averaged Keldysh path integral is given by
\begin{equation}
\overline{Z} = \int \mathcal{D}\left[\chi,\psi\right]
e^{iS},
\end{equation}
where the action reads
\begin{widetext}
\begin{multline}
iS=\frac{i}{2}\int d\tau \sum_i \chi_i \partial_\tau \chi_i - \mu  \int d\tau \sum_m \psi_a^m \psi_b^m - \frac{J^2}{8N^3}  \int d\tau_1 d\tau_2 \left(\sum_i \chi_i(\tau_1)\chi_i(\tau_2) \right)^4 \\
+\frac{K^2}{N}\sum_\alpha \int d\tau_1 d\tau_2 \left(\sum_m \psi^m_\alpha(\tau_1)\psi^m_\alpha(\tau_2) \right) \left(\sum_i \chi_i(\tau_2)\chi_i(\tau_1) \right).
\end{multline}
\noindent Writing the contour integral $\int d\tau$ in terms of forward and backward real time branches we find
\begin{multline}
iS=\sum_{ss'}\int dt dt' \Bigg\{ \frac{i}{2}\sum_i \chi^i_{st} [\sigma^z_{ss'}\delta_{tt'}i\partial_t] \chi^i_{s't'} -  \mu\sigma^z_{ss'} \sum_m \delta_{tt'}\psi_{ast}^m \psi_{bs't'}^m  \\ +ss' \frac{K^2}{N}\sum_\alpha \left(\sum_m \psi^m_{\alpha s t}\psi^m_{\alpha s' t'} \right) \left(\sum_i \chi^i_{s't'} \chi^i_{st} \right)  -  ss'  \frac{J^2}{8N^3}\left(\sum_i \chi_{ist}\chi_{is't'} \right)^4 
 \\
+\frac{N}{2} \Sigma_{ss'}^{tt'}\left(G_{s's}^{t't}-\frac{i}{N}\sum_i  \chi^i_{st}\chi^i_{s't'} \right) + M (\Omega^{\alpha \beta})_{ss'}^{tt'}\left((F_{\beta \alpha})_{s's}^{t't}-\frac{i}{2M}\sum_m \psi^m_{\alpha s t}\psi^m_{\beta s' t'} \right) \Bigg\}
\end{multline}
where we introduced bosonic path integral identities $\int \mathcal{D}[G,\Sigma] \exp{\left(\frac{N}{2}\sum_{ss'}\int \int dt dt' \Sigma^{tt'}_{ss'}\left[G_{s's}^{t't}-\frac{i}{N}\sum_i  \chi^i_{st}\chi^i_{s't'} \right]\right)  }=1$ and $\int \mathcal{D}[\mathcal{G},\Omega] \exp{\left(M\sum_{ss'}\int \int dt dt' (\Omega^{\alpha \beta})_{ss'}^{tt'}\left[(F_{\beta \alpha})_{s's}^{t't}-\frac{i}{2M}\sum_m \psi^m_{\alpha s t}\psi^m_{\beta s' t'}  \right]\right)}=1$.
 As in the Matsubara case, $\Sigma$ and $\Omega$ act as Lagrange multipliers and can be though of as self energies. Gaussian integration over the fermionic degrees of freedom yields
\begin{multline*}
iS = \frac{N}{2}\text{tr}\log A + M\text{tr}\log B \\  +\sum_{ss'}\int dt dt' \left\{\frac{N}{2}\Sigma_{ss'}^{tt'}G_{s's}^{t't} + M [\Omega^{\alpha \beta}]_{ss'}^{tt'} [F_{ \beta \alpha}]_{s's}^{t't}  - \frac{J^2}{8} ss' N (G_{s's}^{t't})^4 + 
2K^2M ss'[F_{\alpha \alpha}]_{s's}^{t't}G_{ss'}^{tt'}
\right\}
\end{multline*}
\end{widetext}
where we defined 
\begin{align}
A_{st,s't'}&=-i^2\left(\sigma^z_{ss'}\delta(t-t')i\partial_t - \Sigma_{ss'}^{tt'} \right) \\ B_{\alpha s t,\beta s' t'}&=i\left([\Omega^{\alpha \beta}]_{ss'}^{tt'} + \mu \sigma^z_{ss'} \sigma^y_{\alpha \beta} \delta(t-t') \right).
\end{align}
The saddle point equations are obtained after following functional derivatives and Fourier transforms
\begin{align}
\frac{\delta S}{\delta \Sigma_{s_1s_2}^{t_1t_2}} =  0, \hspace{20pt}\frac{\delta S}{\delta [\Omega^{\alpha \beta }]_{s_1s_2}^{t_1t_2}} =  0,  \label{delg}\\
\frac{\delta S}{\delta G_{s_1s_2}^{t_1t_2}} =  0,  \hspace{20pt} \frac{\delta S}{\delta  [F_{\alpha \beta }]_{s_1s_2}^{t_1t_2}} =  0. \label{delsigma}
\end{align}
The following then must be valid at saddle point
\begin{align}
\sum_{s_1,\gamma}F^{ \alpha \gamma}_{s s_1}(\omega)\left(\Omega^{\gamma \beta}_{s_1s'}(\omega) + \mu \sigma^z_{s_1s'} \sigma^y_{\gamma \beta}  \right)+\delta_{\alpha \beta}^{ss'} &= 0\label{sp1} \\
G_{s s_1}(\omega)(\omega \sigma_{s_1s'}^z - \Sigma_{s_1s'}(\omega))&=\delta_{ss'}\label{sp2} \\
\frac{1}{2}\Sigma_{ss'}^{tt'}  -\frac{J^2}{2} ss' (G_{s's}^{t't})^3 + 
2K^2 \frac{M}{N} ss'[F_{\alpha \alpha}]_{ss'}^{tt'}&=0 \label{sp3} \\
[\Omega^{\alpha \beta}]_{ss'}^{tt'}  + 
 2K^2 ss'G_{ss'}^{tt'} \delta_{\alpha \beta}
&=0 \label{sp4}
\end{align}
Equation \eqref{sp4} tells us that $\Omega^{\alpha\beta}_{ss'} = \delta^{\alpha\beta}\Omega_{ss'}$ is diagonal. It follows that $F=F^{aa}=F^{bb}$, similar to Matsubara case in Appendix A. Equation \eqref{sp1} can then be recast as
\begin{equation}\label{meq1}
F(\mu^2-\Omega \sigma^z \Omega \sigma^z)=\sigma^z \Omega \sigma^z
\end{equation} where $F$, $\Omega$, $\sigma^z$ are now matrices in Keldysh forward-backward indices.
In this matrix language, we similarly see that equations  (\ref{sp2}-\ref{sp4}) become
\begin{gather}
G(\omega \sigma^z - \Sigma)=I \\
\Sigma = \Sigma_J - 8K^2 p \sigma^z F \sigma^z, \\
\Omega + 2K^2\sigma^z G \sigma^z = 0, \label{meq4}
\end{gather}
where $[\Sigma_J(t-t')]_{ss'}=J^2 ss' (G_{s's}^{t't})^3$ and $p=M/N$. Eliminating $\Omega$ and $F$ by combining equations (\ref{meq1}-\ref{meq4}) we obtain the simplified saddle point equations 
\begin{gather}
    \Sigma = 4p K^2\sigma^z G [\sigma^z(4p-K^2(\sigma^z G)^2)]^{-1} + \Sigma_J \label{seom1}, \\
    G=(\omega \sigma^z - \Sigma)^{-1} \label{seom2}.
\end{gather}
For the non-interacting case $J=0$, we set $\Sigma_J=0$ and combine these two equations to obtain equation \eqref{keldyshnoninteract}. Since we are interested in the equilibrium state we must also impose the fluctuation-dissipation relation to set the temperature $\beta^{-1}$ of the system
\begin{align}\label{eq:FDT}
G^{K}(\omega) = 2i \tanh \left( \frac{\beta (\omega)}{2} \right) \text{Im} G^{R}(\omega).    
\end{align}
In the Keldysh formalism the Green's function $G$ has the following matrix structure\cite{Kamenev2009}
\begin{equation}\label{gmatrixform}
G = \begin{pmatrix}
      G^T     & G^<   \\
    G^>      &  G^{\Tilde{T}}
\end{pmatrix}
=\begin{pmatrix}
    G_{++} & G_{+-}   \\
    G_{-+} & G_{--}
\end{pmatrix}
\end{equation} 
where $G^T$ and $G^{\Tilde{T}}$ are the time ordered and anti-time ordered Green's functions, respectively. $G^<$ and $G^>$ are the lesser and the greater Green's functions. These four quantities are not independent; by construction, they are related by $G^T + G^{\Tilde{T}} = G^{<} + G^{>}$. Their relation to the Keldysh Green's function is given by $G^K=G^{<}+G^{>}$ while the retarded Green's function $G^R$ is given by $G^R= G^T- G^{<}$ from which we can obtain the spectral functions $A=-2\text{Im}G^R$.

\subsection{Numerical Solution} 
\noindent We solve the Keldysh saddle point equations (\ref{seom1},\ref{seom2}) iteratively using a discrete real-time/frequency array of matrices of the form \eqref{gmatrixform} for the Green's functions $G$. Non-interacting equations of motion \eqref{keldyshnoninteract} can be solved by direct iteration in real frequency space. However, the interacting case $J>0$, requires switching between real time and frequency representations at each step of iteration as described in the following. Starting with an ansatz for $G[i]$ where $i$ is the iteration index, we first compute $[\Sigma_J(t)]_{ss'}=J^2 ss' (G_{s's}(-t))^3$ after inverse Fourier transforming $G(\omega)$. We next Fourier transform $\Sigma_J(t)$ to substitute in \eqref{seom1} and compute the total self energy $\Sigma[i]$ in the frequency representation. We then mix the new Green's function which we compute using $\Sigma[i]$ in Eq.\ \eqref{seom2}, with the one from the previous iteration $G[i]$ according to prescription 
\begin{equation}
G[i+1]=\alpha\frac{1}{\omega \sigma^z - \Sigma[i]} + (1-\alpha)G[i],
\end{equation} 
where $\alpha$ is the mixing parameter. Fast convergence is achieved for $\alpha=0.2$ which we use in all calculations in this work. We use FFT algorithms \cite{scipy} for shorter computation times and repeat this iterative procedure until the solution converges. Since we are interested in equilibrium, we also constrain each iteration with the fluctuation-dissipation relation \eqref{eq:FDT} to fix the temperature of the system.
\end{document}